\documentstyle[12pt,aaspp4,flushrt]{article}
 
\input epsf

\newcommand{\al}{\alpha}

\newcommand{\vs}{{v_{{\scriptscriptstyle 12}}}}

\newcommand{\be}{\begin{equation}}
\newcommand{\ee}{\end{equation}}
\newcommand{\gsim}{\stackrel{>}{\sim}}
\newcommand{\lsim}{\stackrel{<}{\sim}}
\newcommand{\bea}{\begin{eqnarray}}
\newcommand{\eea}{\end{eqnarray}}
\newcommand{\bean}{\begin{eqnarray*}}
\newcommand{\eean}{\end{eqnarray*}}
\newcommand{\dd}{\partial}
\newcommand{\xb}{\bar{\xi}}
\newcommand{\xbb}{\bar{\hspace{-0.08cm}\bar{\xi}}}

\begin{document}
\tighten

\title{A Simple Method for Computing the Non-Linear Mass 
Correlation Function with Implications for Stable Clustering}
\author{Robert R. Caldwell\altaffilmark{1,5},
Roman Juszkiewicz\altaffilmark{1-5}, \\
Paul J. Steinhardt\altaffilmark{1,5},
and \hbox{Fran{\c c}ois R. Bouchet\altaffilmark{4,5}}}

\altaffiltext{1}{Department of Physics, Princeton University,
Princeton, NJ 08544}
\altaffiltext{2}
{D{\'e}partement de Physique Th{\'e}orique, Universit{\'e}
de Gen{\`e}ve, CH-1211 Gen{\`e}ve, Switzerland}
\altaffiltext{3}
{Copernicus Astronomical Center, 00-716 Warsaw, Poland}
\altaffiltext{4}
{Institut d'Astrophysique de Paris, CNRS, Paris, France}
\altaffiltext{5}
{Isaac Newton Institute for Mathematical Sciences,
Cambridge, England CB3 0EH}

\begin{abstract} 

We propose a simple and accurate method for computing analytically the  mass
correlation function for cold dark matter and  scale-free models that fits
N-body simulations over a range that extends from the linear to the strongly
non-linear regime. The method, based on the dynamical evolution of the pair
conservation equation, relies on a universal relation between the pair-wise
velocity and the smoothed correlation function valid for high and low density
models, as derived empirically from N-body simulations. An intriguing
alternative relation, based on the stable-clustering hypothesis, predicts a
power-law behavior of the mass correlation function that disagrees with N-body
simulations but conforms well to the observed galaxy correlation function if
negligible bias is assumed. The method is a useful tool for rapidly exploring a
wide span of models and, at the same time, raises new questions about large
scale structure formation.

\end{abstract}


\tighten

%%%%%%%%%%%%%%%%%%%%%%%%%%%%%%%%%%%%%%%%%%%%%%%%%%%%%%%%%%%%%%%%%%%%%%%%%%%%%
 
\eject

Understanding the origin and evolution of the clustering pattern of galaxies is
one of the most important goals of cosmology. Until now, this problem has been
investigated using a four-fold path: (1) perturbation theory (for a review of
recent advances, see  \cite{rj96} and references therein); (2) a kinetic
description, adapted from the BBGKY hierarchy, used in plasma physics 
(\cite{jp80},  \S IV); (3) N-body simulations (e.g., \cite{aj98}, hereafter
VIRGO); (4) and semi-analytic fits to N-body results, based on the so-called
universal scaling hypothesis (see \cite{ah91,jmw95,pd96,Ma98}). The advantages
and limitations of these methods are often complementary. For example, 
applying perturbation theory often leads to analytic results for a wide class
of models while the N-body simulations allow a study of only one model at a
time. On the other hand, perturbation theory works only in the weakly
non-linear regime while N-body experiments describe the fully non-linear
dynamics, albeit over a limited dynamical range. The subject of the present
study is an analytic ansatz for the evolution of the two-point correlation
function of density fluctuations spanning the linear and non-linear regime,
which builds on all four methods described above.   

Our approach is based on the pair conservation equation, which relates the mean
(pair-weighted) relative velocity of a pair of particles to the time evolution
of the correlation function in a self-gravitating gas:
\begin{equation} {a\over 3[1+\xi(x,a)]} \, {\dd\xb(x,a)\over \dd a} \; = \; -
\,  {\vs(x,a)\over Hr}~,  \label{full} 
\end{equation} 
where $a(t)$ is the expansion factor with $a=1$ at present,  $r = ax$ is the
proper separation and $H(a)$ is the Hubble parameter (see \cite{md77,jp80}).
Here $\xb(x,a)$ represents the two-point correlation function averaged over a
ball of comoving radius $x$:  
\begin{equation} \xb(x,a) \; =  \; {3 \over x^3} 
\, \int_0^x \xi(y,a)y^2 dy \; . \label{xb} 
\end{equation}  
The approximate solution  of (\ref{full}) is known in the large  separation
limit, where $|\xi|\ll1$ (linear regime); the stable clustering hypothesis is
often invoked to describe the small separation limit, where $\xi \gg 1$
(non-linear regime).  Hence, equation (\ref{full}) is ``a guide to speculation
on the behavior of the correlation function'' (\cite{jp80}, p.268) since an
assumed $\vs$ implies a function $\xi$ that should agree with the weak and
strong field limits, and interpolates between these limits in a reasonable way.

An approximate universal relation between the pair-wise velocity and the
smoothed correlation function has been conjectured by \cite{ah91} on the
basis of N-body simulation results and further explored in  \cite{np94} and
\cite{pe98}. In the past, the relation has been used in attempts to derive a
general functional that converts directly from a linear to a non-linear mass
correlation. In this paper, we present a simple extension of the relation that
applies to both high- and low-density models, but take a different
approach to obtaining the non-linear correlation function. Namely, we use the
universal relation to close (\ref{full}); we then evolve the resulting
partial differential equation to compute the non-linear correlation function. 
This turns out to be a fast and surprisingly accurate method that matches
N-body results for a wide variety of cold dark  matter (CDM) models. As a
stand-alone computer program, the algorithm can be adopted on a programmable
calculator; or, it can easily be incorporated in more sophisticated programs
that predict other cosmological properties, such as CMBFAST (\cite{CMBFAST}).

\begin{figure*}
\plotone{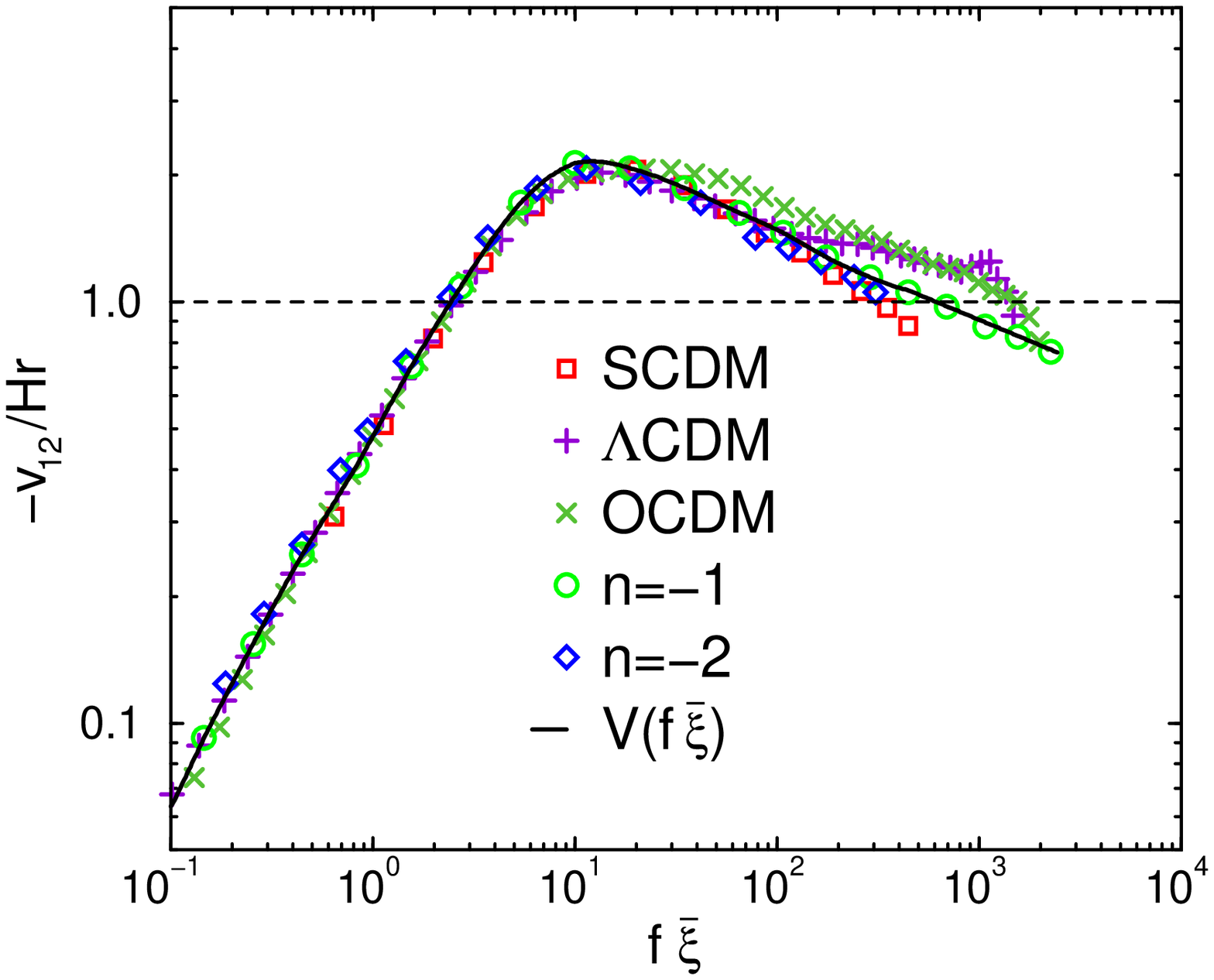}
\caption{The pair-wise velocity in units of the Hubble velocity, $-\vs/Hr$ is
shown as a function of the smoothed correlation function times the linear
density growth factor, $f\,\xb$ as determined by N-body simulations (scale free
and SCDM from \protect{\cite{jain97}}, Figures 3 \& 8; $\Lambda$ and OCDM from
VIRGO).  Not only for these examples, but for a wider variety of initial
conditions and at different times, the pair-wise velocity displays nearly the
same shape in $\xb$, leading us to a nearly universal relation, $V[f\xb] \equiv
-\vs/Hr$. \label{figure1}}
\end{figure*}

\begin{figure}
\plotone{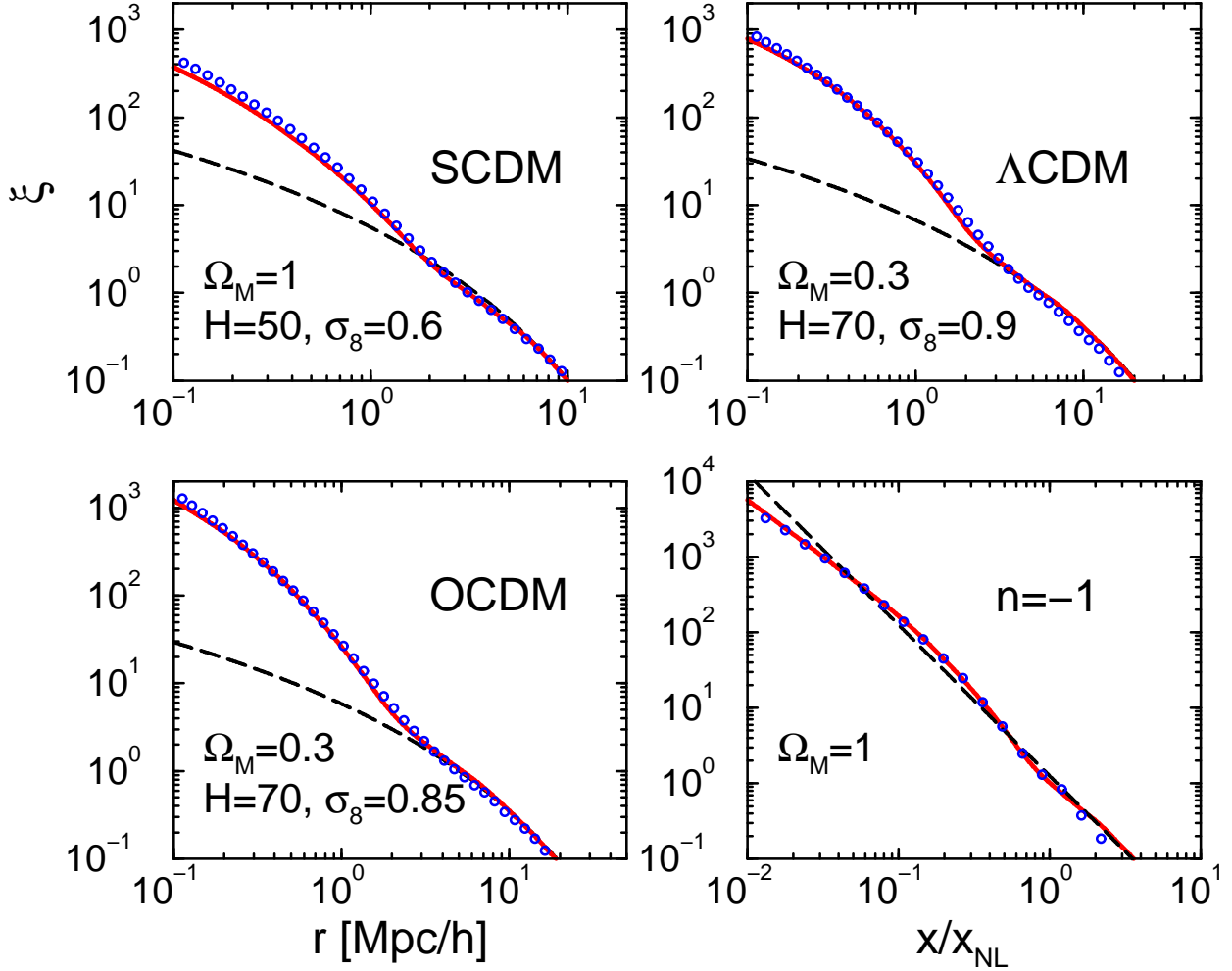}
\caption{The non-linear correlation function $\xi$ as a function of separation
is shown for various cases. The non-linear $\xi$ obtained from our method
(solid lines) is in excellent agreement with the N-body  results (circles). The
dashed line is the linear $\xi$. The cosmological models are from the VIRGO
simulation, whereas the $n=-1$ scale-free model is from
\protect{\cite{jain97}}, Figure 1. For $n=-1$, $x=x_{NL}$ at $\xi=1$. 
\label{figure2}}
\end{figure}

Figure \ref{figure1} clearly illustrates the nearly model-independent
relationship between the pair-wise velocity and the smoothed correlation
function, observed in N-body simulations for a wide range of perturbation
spectra.  We define this relation as
\begin{equation}
V[\,f\xb\;] \equiv -{\vs \over Hr}.
\end{equation}
Compared to \cite{ah91}, a novel feature is plotting the relation in terms of
$f(\Omega) \bar{\xi}$, where $f \equiv d \ln D/d \ln a$ is the  standard linear
density growth factor,  rather than $\bar{\xi}$ alone --- a difference that is
essential for extending the relation to low-density models. The evolved,
non-linear clustering of scale-free spectra with $n=-1,\,-2$ as well as the
CDM  family of models produces a very similar relation between $-\vs/Hr$ and
$\xb$. An excellent fit to the functional relation $V[f\xb(x,a)]$ in Figure
\ref{figure1}, based on the $n=-1$ curve,  is given by
\begin{equation}
V[x] = \cases{{2 \over 3}\, x & $x < 0.15$ \cr
0.7 \, x \,\exp(-0.31 \, x^{0.61}) & $0.15 \le x < 20$ \cr
3.3 \, x^{-0.17} & $x \ge 20$ }
\label{fitfunc}
\end{equation}
valid for $x \lsim 10^3$.
In this paper, we use this fitting formula, designed for  the $n=-1$ curve, as
the expression for $V[x]$ in (\ref{full}) to be applied to all models.   We
find this to be sufficient to reproduce N-body results to within  10\% accuracy
over the range of models and scales shown in Figure \ref{figure2}, which
extends deep into the non-linear regime.    To push to  lower density models
and improve the accuracy further,  it would be simple to modify the algorithm 
for $V[x]$ to include the $\Omega$-dependent rise near $x\gsim 20$.

Starting with the linear correlation function, and armed only with information
about the background cosmological evolution, we propose to dynamically obtain
the non-linear $\xi$ as a function of separation and time. Here then is our
idealized procedure in three steps.

{\it 1. Reformulate:} We first rewrite the partial differential equation as
\begin{equation}
{\dd\ln\xb\over \dd \ln a} \; = \; 3 \,{(1 + \xi) \over \xb } \, V[ f \xb]  
\label{newpde}
\end{equation}
where $V[x]$ is given in equation (\ref{fitfunc}).
 
{\it 2. Initialize:} The initial conditions are set by the linear correlation
function at a red shift $z=-1 + 1/a_i$ such that $\xi(x,a_i)$ is less than
unity for all $x$ of interest. Our procedure assumes that only the amplitude,
and not the shape of the linear correlation function has changed over this
interval, as occurs for cold dark matter models with a primordial spectrum of
adiabatic density perturbations. Hence, this procedure will not apply to
cosmological models with a late-time decaying neutrino, but will apply to hot
dark matter models wherein the shape of the linear power spectrum is set by red
shift $z\sim100$. 

{\it 3. Evolve:} We numerically solve the partial differential equation and
evolve $\xb$. At each step in the evolution, we use $\xi = \xb \times (1 -
\bar\gamma/3)$ with $\bar\gamma\equiv -d\ln\xb/d\ln{x}$ to determine the
correlation function. The value of $f$ is updated at each step in $a$ as
appropriate for the cosmology.

The remarkable results are shown in Figure \ref{figure2}.  Here we see that
our  simple procedure gives excellent agreement with N-body simulations. Based
on the span of behavior in the cosmic time evolution and the shape of
correlation function, we expect this procedure should be valid for a wide range
of cosmological models, including quintessence (\cite{CDS}) and models with
tilted spectra.

Figure \ref{figure2} demonstrates that we have obtained a simple and powerful
new tool for rapidly and accurately obtaining the non-linear power spectrum for
a wide range of models.  However, the physical origin of the nearly
model-independent relation $V[f\xb]$ is not understood in detail.  In the
linear regime, perturbation theory predicts $-\vs/Hr = (2/3) f \xb$.  In the
non-linear regime, Padmanabhan {\it et al.} (1996) have suggested that insight
may be obtained by comparison to the case of the gravitational collapse of a
spherical top hat mass distribution.  Using their solution (eqs. (16-19) in
their paper),  we find that  $-\vs/Hr$ is linearly proportional to
$f\bar{\xi}$  times a slowly decreasing function of $\xb$ for a surprisingly
wide range of $\bar{\xi} \gg 1$, including $\bar{\xi}= 10$,  the turnover point
in  Figure \ref{figure1}. In particular, for $\bar{\xi}=10$, $-\vs/Hr = 1.77 \,
f(\Omega)$, which is similar to  $-\vs/Hr \approx 2  f(\Omega)$ in Figure
\ref{figure1}.

In the strongly non-linear regime, $\bar{\xi} \gg 10$,  Figure \ref{figure1}
shows  a visible difference in the shape of $V[f\xb]$ between the high density
and low density models. This may be due to the suppression of linear growth,
which occurs at late times in low density models and leads to the enhanced
clustering on small scales relative to large scales. However, this has a
negligible effect on the computed non-linear correlation function. For example,
using the curve for $\Lambda$CDM shown in Figure \ref{figure1} as the basis of
$V$ in our procedure, we find the amplitude of the non-linear correlation
function differs by only $10\%$ at $r\sim 0.1$ Mpc/h.  For the models shown,
this accuracy is comparable to  what is obtained by \cite{ah91} and
\cite{pd96}.   The advantage here is that our method can be immediately applied
to new types of CDM models ({\it e.g.}, quintessence cosmologies) without
having to run new N-body simulations to fix fitting parameters.

An important issue raised by our ansatz is the validity of the stable
clustering hypothesis. The stable clustering regime corresponds to the limit
where particle pairs detach from the  Hubble flow and $-\vs/Hr \to 1$. Figure
\ref{figure1} shows that  $-\vs/Hr$ first  overshoots unity by a factor of two
and then rebounds towards unity.  However, it is not clear whether it 
converges to unity at $\xb \approx 1000$ or possibly oscillates if the
simulations are extended to higher values of $\xb$.    

\begin{figure}
\plotone{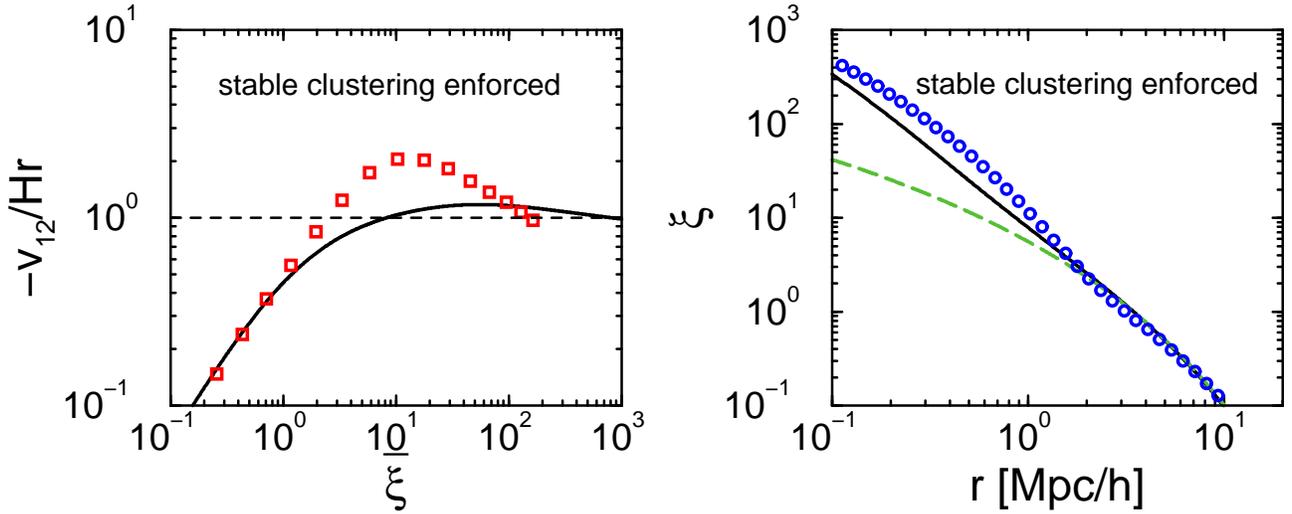}
\caption{The pair-wise velocity and non-linear correlation function are shown  
for a case (solid curves) in which we have forced a rapid convergence to  
stable clustering, $-\vs/Hr = 1$. For the purposes of illustration,  we have
used the JSD ansatz  described by equation (\protect{\ref{2nd}}), although
substituting any similar shape for $V[f\bar{\xi}]$   would produce a similar 
effect on the mass correlation function.  The rapid approach to stable
clustering and the absence of any significant overshoot, as seen in the solid
curve on the left,  results in the power law behavior $\xi\propto r^{-1.7}$ on
the right.  The circles and squares in each panel are the N-body results for
SCDM. \label{figure3}}
\end{figure}

It is interesting to compare the predictions of our ansatz if the relation
between $-\vs/Hr$ and $\bar{\xi}$ is modified to enforce more rapid convergence
to stable clustering. A ready example is an alternative ansatz based on the
pair conservation equation  recently proposed by \cite{rj98} (hereafter JSD). 
Their ansatz for $\vs$, based on an interpolation between the behavior
predicted by perturbation theory in the weakly non-linear regime and stable
clustering in the strongly non-linear regime is given by
\begin{equation}
\vs(x,a)  \; := \;
- \, {\textstyle {2\over 3}}\,Hrf\,
\xbb(x,a)\,\left[\,1 \; + \;\al\;\xbb(x,a) \; \right] \label{2nd}  
\end{equation}
where $\, \xbb  \; \equiv \; \xb/(1 + \xi)$ and $\alpha$ is a function  which
controls the strength of the non-linear feedback. Here we use $\alpha=1.8 -
1.1\gamma$, based on perturbation theory (see \cite{rs96}), where $\gamma$ is
the slope of the correlation function at $\xi=1$. The key point, as shown in
Figure \ref{figure3}, is that the pair-wise velocity rapidly approaches the
stable clustering limit by $\xb\sim 10$, and remains there to within $\sim
20\%$  on smaller scales in the more strongly non-linear regime. This means
particle pairs separated by $\lesssim 1$ Mpc/h have the rough behavior of
virialized objects, such as clusters and galaxies. 

The correlation function $\xi$ obtained by closing the pair conservation
equation with (\ref{2nd}), as shown in   Figure \ref{figure3}, displays a
power-law  behavior in the non-linear regime with index $\sim -1.7$, in
disagreement with N-body simulations of CDM but  curiously  similar to  the
galaxy correlation function observed in the APM survey (\cite{sm96}). This
result is not unique to the JSD ansatz;  substituting any shape similar to that
shown in the top panel of Figure \ref{figure3} for $V[f\bar{\xi}]$ into our
ansatz would produce a similar effect on the mass correlation function. In the
past, the disagreement between the dark matter power spectrum observed in
simulations, which shows no evidence of power-law behavior, and the simple
power-law observed in the galaxy correlation function has been attributed to a
scale-dependent bias.  The conventional picture is that the bias function,
$b(r)$, is just so as to cause a non-power-law behavior in the dark matter to
be translated into a power-law behavior of the galaxy correlation function,
$\xi_g(r) = \xi(r)b^2(r)$. 

Our present findings concerning the dependence of the mass correlation function
on the model independent  $-\vs/Hr$ vs. $f\, \bar{\xi}$ relation suggests a
radical possibility.  Perhaps  the bias factor is completely negligible and, 
instead,  CDM models are missing some  important   physical feature ({\it e.g.}
mechanics of  galaxy formation, or some property of the dark matter) which
causes rapid convergence to  stable clustering in the  non-linear regime
($-\vs/Hr \rightarrow 1$ without substantial  overshoot) and to power-law
behavior of the correlation function.  The ansatz illustrated in Figure
\ref{figure3}, which enforces stable clustering by {\it fiat}, may be
implicitly describing a  modified CDM model of galaxy formation  which 
incorporates the new physical feature.   The difference between N-body
simulations and observations, whether due to bias in the conventional picture
or to something more radical, such as a modification to CDM, can perhaps be
determined empirically by studying red shift distortion on the scales where
N-body simulation suggests overshoot in  $-\vs/Hr$ and the ansatz of Figure
\ref{figure3} does not.  

In sum, our studies have produced a  simple recipe for computing the non-linear
power spectrum for a wide range of models.  (Upon publication of the paper, we
will make a program available at feynman.princeton.edu/$\sim$steinh.)  A key
feature of the method is  the universal function relating the pair velocity to
the mass correlation function  which does not converge rapidly to the stable
clustering limit, but, rather,  overshoots by a factor of two.  This feature is
responsible for the fact that the mass correlation function does not approach a
power-law. Our studies have also raised several  interesting issues in
structure formation: why is $-\vs/Hr$ vs. $f \,\xb$ so similar for a wide range
of models?  can the universal relation be derived from theory? for what
range of models is the relation  model-independent?  how does the universal
relation ultimately approach the stable clustering limit at small scales (if it
does at all)? and, does the success of a universal relation based on the stable
clustering hypothesis, as in Figure \ref{figure3}, suggest a viable,
alternative explanation for the power-law behavior of the galaxy correlation
function?

%%%%%%%%%%%%%%%%%%%%%%%%%%%%%%%%%%%%%%%%%%%%%%%%%%%%%%%%%%%%%%%%%%%%%%%%%%%%%
\acknowledgements

We would like to thank Dick Bond, Marc Davis, Josh Frieman, Andrew Hamilton, 
Chung-Pei Ma, Jim Peebles,  David Spergel, and  Roman Scoccimarro for useful
conversations. We also thank  Bhuvnesh Jain  and Volker Springel and the VIRGO
collaboration for providing N-body simulation results. This work was carried
out, in part, at the Isaac Newton Institute for Mathematical Sciences during
their program on Structure Formation in the Universe. We would like to thank
the organizers, N. Turok and V. Rubakov, and the staff of the Institute for
their support and kind hospitality.

The work of RRC and PJS was supported by the US Department of Energy grant
DE-FG02-91ER40671. The work of RJ was supported by grants from the Polish
Government (KBN grants  No. 2.P03D.008.13 and 2.P03D.004.13), the Tomalla
Foundation of Switzerland, by the Poland-US M. Sk{\l}odowska-Curie Fund. FRB
and RJ were supported by the Franco-Polish collaboration grant (Jumelage).

\vfill
\eject

%%%%%%%%%%%%%%%%%%%%%%%%%%%%%%%%%%%%%%%%%%%%%%%%%%%%%%%%%%%%%%%%%%%%%%%%%%%%%

%%%%%%%%%%%%%%%%%%%%%%%%%%%%%%%%%%%%%%%%%%%%%%%%%%%%%%%%%%%%%%%%%%%%%%%%%%%%%

%%%%%%%%%%%%%%%%%%%%%%%%%%%%%%%%%%%%%%%%%%%%%%%%%%%%%%%%%%%%%%%%%%%%%%%%%%%%%
%       Figure 1
%%%%%%%%%%%%%%%%%%%%%%%%%%%%%%%%%%%%%%%%%%%%%%%%%%%%%%%%%%%%%%%%%%%%%%%%%%%%%
%%%%%%%%%%%%%%%%%%%%%%%%%%%%%%%%%%%%%%%%%%%%%%%%%%%%%%%%%%%%%%%%%%%%%%%%%%%%%

%%%%%%%%%%%%%%%%%%%%%%%%%%%%%%%%%%%%%%%%%%%%%%%%%%%%%%%%%%%%%%%%%%%%%%%%%%%%%
%       Figure 2
%%%%%%%%%%%%%%%%%%%%%%%%%%%%%%%%%%%%%%%%%%%%%%%%%%%%%%%%%%%%%%%%%%%%%%%%%%%%%
%%%%%%%%%%%%%%%%%%%%%%%%%%%%%%%%%%%%%%%%%%%%%%%%%%%%%%%%%%%%%%%%%%%%%%%%%%%%%

%%%%%%%%%%%%%%%%%%%%%%%%%%%%%%%%%%%%%%%%%%%%%%%%%%%%%%%%%%%%%%%%%%%%%%%%%%%%%
%       Figure 3
%%%%%%%%%%%%%%%%%%%%%%%%%%%%%%%%%%%%%%%%%%%%%%%%%%%%%%%%%%%%%%%%%%%%%%%%%%%%%
%%%%%%%%%%%%%%%%%%%%%%%%%%%%%%%%%%%%%%%%%%%%%%%%%%%%%%%%%%%%%%%%%%%%%%%%%%%%%


\begin{thebibliography}{}
 
\bibitem [Caldwell, Dave, \& Steinhardt 1998]{CDS}
Caldwell, R.R., Dave, R., \& Steinhardt, P.J. 1998, \prl, 80, 1586.

\bibitem [Davis \& Peebles 1977] {md77}
Davis, M. \& Peebles, P.J.E 1977, \apjs, 34, 425.

\bibitem [Hamilton {\it et al.} 1991] {ah91}
Hamilton, A.J.S., Kumar, P., Lu, E., \& Matthews, A. 1991,
\apjl, 374, L1.

\bibitem [Jain {\it et al.} 1995] {jmw95}
Jain, B., Mo, H.J., \& White, S.D.M. 1995, \mnras, 276, L25.

\bibitem [Jain 1997]{jain97}
Jain, B. 1997, \mnras, 287, 687.

\bibitem [Jenkins {\it et al.} 1998] {aj98} 
Jenkins, A., et al. 1998, \apj, 499, 20. (VIRGO)

\bibitem [Juszkiewicz \& Bouchet 1996]{rj96}
Juszkiewicz, R., \& Bouchet, F.R. 1996, in {\it
Proc. XXX Moriond Meeting, Clustering in the Universe},
ed. S. Maurogordato et al. (Editions Frontieres: Paris), p. 167.
 
\bibitem [Juszkiewicz {\it et al.} 1999]{rj98}
Juszkiewicz, R., Springel, V., Durrer, R. 1999, \apj, 518, L25. (JSD)

\bibitem [Ma 1998]{Ma98}
Ma, C.-P. 1998, \apj, 508, L5.

\bibitem[Maddox {\it et al.} 1996]{sm96}
Maddox, S.J., Efstathiou, G., \& Sutherland, W.J.,  1996, \mnras,
283, 1227.

\bibitem[Nityananda \& Padmanabhan 1994]{np94}
Nityananda, R. \& Padmanabhan, T. 1994, \mnras, 271, 976. 

\bibitem[Padmanabhan {\it et al.} 1996]{pcos96}
Padmanabhan, T., Cen, R., Ostriker, J.P., \& Summers, F.J. 1996, 
\apj, 466, 604.

\bibitem[Padmanabhan \& Engineer 1998]{pe98}
Padmanabhan, T. \& Engineer, S. 1998, \apj, 493, 509.

\bibitem [Peacock \& Dodds 1996] {pd96}
Peacock, J.A., \& Dodds, S.J. 1996, \mnras, 280, L19.

\bibitem [Peebles 1980] {jp80}
Peebles, P.J.E. 1980, The Large-Scale Structure of the Universe,
(Princeton:  Princeton University Press) 

\bibitem [Scoccimarro \& Frieman 1996] {rs96}
Scoccimarro, R., \& Frieman, J. 1996, \apj, 473, 620.

\bibitem[Seljak \& Zaldarriaga 1996]{CMBFAST}
Seljak, U. \& Zaldarriaga, M. 1996, \apj, 469, 437.

\end{thebibliography}
\end{document}